# When does deep learning fail and how to tackle it? A critical analysis on polymer sequence-property surrogate models

Himanshu and Tarak K Patra[*]

Department of Chemical Engineering, Center for Atomistic Modeling and Materials Design and Center for Carbon Capture Utilization and Storage, Indian Institute of Technology Madras, Chennai TN 600036, India

**Abstract:**

Deep learning models are gaining popularity and potency in predicting polymer properties. These models can be built using pre-existing data and are useful for the rapid prediction of polymer properties. However, the performance of a deep learning model is intricately connected to its topology and the volume of training data. There is no facile protocol available to select a deep learning architecture, and there is a lack of a large volume of homogeneous sequence-property data of polymers. These two factors are the primary bottleneck for the efficient development of deep learning models. Here we assess the severity of these factors and propose new algorithms to address them. We show that a linear layer-by-layer expansion of a neural network can help in identifying the best neural network topology for a given problem. Moreover, we map the discrete sequence space of a polymer to a continuous one-dimensional latent space using a machine learning pipeline to identify minimal data points for building a universal deep learning model. We implement these approaches for three representative cases of building sequence-property surrogate models, viz., the single-molecule radius of gyration of a copolymer, adhesive free energy of a copolymer, and copolymer compatibilizer, demonstrating the generality of the proposed strategies. This work establishes efficient methods for building universal deep learning models with minimal data and hyperparameters for predicting sequence-defined properties of polymers.

[*]Author to Correspond, E-mail: tpatra@iitm.ac.in

## I. Introduction

Traditional experiments and computer simulations are limited by their inability to rapidly measure polymer properties, and, thus, inadequate to screen the astronomically large chemical and conformational space of a polymer. Recent advances in machine learning (ML) and increasing data and software availability trends can address this problem and accelerate polymer design.[1–5] Significant progress has been made to create data-driven models that predict polymer properties. These models are built by collecting candidate polymers and labeling them by their properties, which are calculated using physics-based methods. A large variety of machine-readable fingerprints and chemical descriptors are developed to represent polymers for ML models.[6–9] The fingerprint-property data is utilized for training and building an ML model. Such an ML model serves as a cheaper, albeit low-fidelity, surrogate for the high-fidelity first-principle-based simulations and experiments that are expensive. There exist a large number of numerical frameworks, such as support vector regression, random forest, and deep neural network (DNN) to build these ML models. Among these, DNN appears to be more versatile and transferable and provides a flexible mathematical framework to model structure-property correlation. DNNs have been progressively used to build structure-property models of a wide range of materials including polymers.[10–19] They consist of a large number of nodes arranged in several intermediate layers between the input and output layers. Some of the important factors that impact the performance of a DNN are weight initialization, activation function of its nodes, learning rate, network topology, stopping criteria, and loss optimization algorithm. Among these, the number of nodes and their arrangement in the intermediate layers plays a key role in determining the accuracy and efficiency of the model. However, there is no systematic guideline to build DNNs that are computationally efficient yet make good-quality predictions. The connection between a DNN topology and the quality of its predictions is not well-established. Moreover, there is no comprehensive understanding of the amount of training data required for building a DNN model that can predict a wide variation in a material's property.

We address the above problems of DNN model development for a representative case of materials properties prediction, viz., sequence-property surrogate model of polymers. The sequence of a polymer appreciably impacts its bulk and single-molecule properties. Glass transition, ion transport, thermal conductivity, a single-molecule radius of gyration, and multimolecular aggregation are all impacted by monomer-to-monomer sequence details of a polymer.[20–26] This sequence-property correlation of a polymer is poorly understood due to its enormous sequence and composition space, and DNNs have been recently used to address this problem and predict sequence-defined properties of polymers.[8,27–29] However, no agreed-upon strategy has emerged to decide the minimum sequence-property data required to build these models. Also, it is not clear what would be the most efficient neural network topology for the sequence-property metamodel of polymer. The primary bottleneck in building a universal model is the astronomically large number of sequences that are possible for a copolymer, and the sequence-specificity is so profound that a subtle change in the copolymer sequence results in a significant change in the properties of interest.[30–32] Oftentimes, the optimal property is present in a non-intuitive, seemingly arbitrary polymer sequence, the sequence-specificity of which is unknown.[20,21,23] Learning and predicting these variations in structure-property relations of a polymer are challenging. There are no analytical methods that can estimate the extremum of a property and the corresponding sequences. It is also challenging to establish the sequence space as a function of a few coordinates. Therefore, building a transferable model remains a substantially complex task.

While the potential of ML predictive models such as DNNs is very lucrative, they are interpolative and, therefore, it is not always clear how one should go about training a neural network to exhaustively fit the entire configurational space of a given system. Currently, DNNs are trained by generating a large quantity of training data in hopes that they have adequately sampled the configurational space of a molecular system. This can, however, be an increasingly prohibitive task when it comes to generating data using computationally expensive physics-based methods. As such, it is desirable to train a model using the absolute minimal data set possible, especially when the costs of high-fidelity calculations are high. In the recent past, we have proposed active learning methods to sample configurational space for collecting DNN training data in the context of neural network potential development.[33–35] Several other active learning strategies, such as QBC (query by committee)[36], DP-GEN (deep potential generator)[37], and adaptive Bayesian inferences[38] for data selection and building transferable neural network models. Moreover, there have been other attempts, such as transfer learnings, to build models with minimal training data. In transfer learning, a model trained on a different property with a given abundant data set is reused and transferred to build another model for a target task with considerably small data.[39,40] All of these require physics-based property calculation while selecting the training data. Therefore, selecting the minimal amount of candidate structures without knowing their properties a priori remains an elusive and attractive goal of ML model development.

The objectives of this work are to build an algorithm to identify the hyperparameters of a DNN, estimate the limitation of a DNN, and, finally, establish a framework to build DNN models that are transferable across the sequence space without the need to generate a large volume of sequence-property data. To accomplish these objectives, we consider three representative problems – the radius of gyration of a copolymer in an infinitely dilute solution, copolymer compatibilizer, and copolymer adsorption on a surface. We propose a systematic linear expansion of DNN architecture to identify the best surrogate models for all three cases. This approach does not require any special optimization algorithm to explore enormously large possibilities of a DNN topology. We use this protocol to develop DNN models that predict sequence-defined properties of polymers with more than 95% accuracy. Secondly, we build a DNN model using training data that represent a specific range of property and test this model's ability to predict the property that is outside the training data. We show that the performance of a DNN declines when the target property is outside the known range of property. We propose a new framework to tackle the transferability problem of ML by leveraging the power of convolution DNN autoencoder that automatically extracts features of a molecular system. We construct a one-dimensional sequence space and sample the sequences uniformly covering the entire space. This collection of points serves as the training data for our DNN model. We show that a model based on ~500 data points, which are selected intelligently, can predict the properties of ~40000 sequences very accurately. We expect this model to predict the properties of all possible sequences of a copolymer, which is ~$10^{30}$ for a binary copolymer of chain length 100. Although the current study focuses on sequence-property ML models, these methods are extensible for other classes of properties and materials. We expect that these new approaches to data and hyperparameter selections will accelerate the progress of ML model development.

## II. Polymer Sequence-Property Data

In this study, we focus on three sequence-defined properties of a binary copolymer, viz., the radius of gyration in an infinitely dilute solution, compatibilization of a polymer blend, and the adsorption-free energy on a patterned surface. The data are collected from recent molecular simulation studies that use the Kremer-Grest bead-spring phenomenological model[41,42] to investigate sequence-property correlations. In this phenomenological model, two chemical moieties are linearly connected to form a copolymer. The interaction parameters of the moieties are adjusted to represent their chemical affinity in a given system. It is a standard and popular model for studying generic polymer properties in molecular simulations without considering specific polymer chemistry and condition. This simple model is computationally very efficient and can be mapped to real polymers by tuning its parameters.[43] The schematic representations of the systems and the distribution of data are shown in Figure 1. The radius of gyration of a copolymer in an implicit solvent is taken from our recent study.[22] In this study, a polymer of chain length *N=100* with an equal composition of both moieties is simulated in an implicit solvent condition, as schematically shown in Figure 1A. A large number of sequences are sampled using a molecular dynamics simulation-based evolutionary algorithm. The data set consists of ~40000 sequences and their radius of gyration. The second data set (*cf.* Figure 1B and E) corresponds to a copolymer compatibilizer.[24] Copolymer compatibilizers are surfactant molecules designed to improve the stability of an interface. They are deployed to enhance material properties in settings ranging from emulsions to polymer blends. A major compatibilization strategy employs block or random copolymers composed of distinct repeat units with preferential affinity for each of the two phases forming the interface, as shown in Figure 1B. In recent studies, we have shown that the surface tension of the interface is very sensitive to the exact sequence of monomers in the copolymer compatibilizer.[23,24] Here we utilize the sequence-surface tension data from these studies to build ML models. The data set corresponds to a blend of two homopolymers, and the compatibilizer is a copolymer of chain length *N=20*. It consists of 1300 sequences and the corresponding surface tension values. Further, we have considered the sequence-defined adsorption-free energy of a copolymer on a decorated surface, as schematically shown in Figure 1C. The data is taken from the work of Shi et al.,[44] wherein they reported the free energy of a copolymer adhesion on a functional surface made of chemically similar particles using molecular simulation. This data set consists of ~20000 sequences of a copolymer of chain length *N=20* and their free energies.

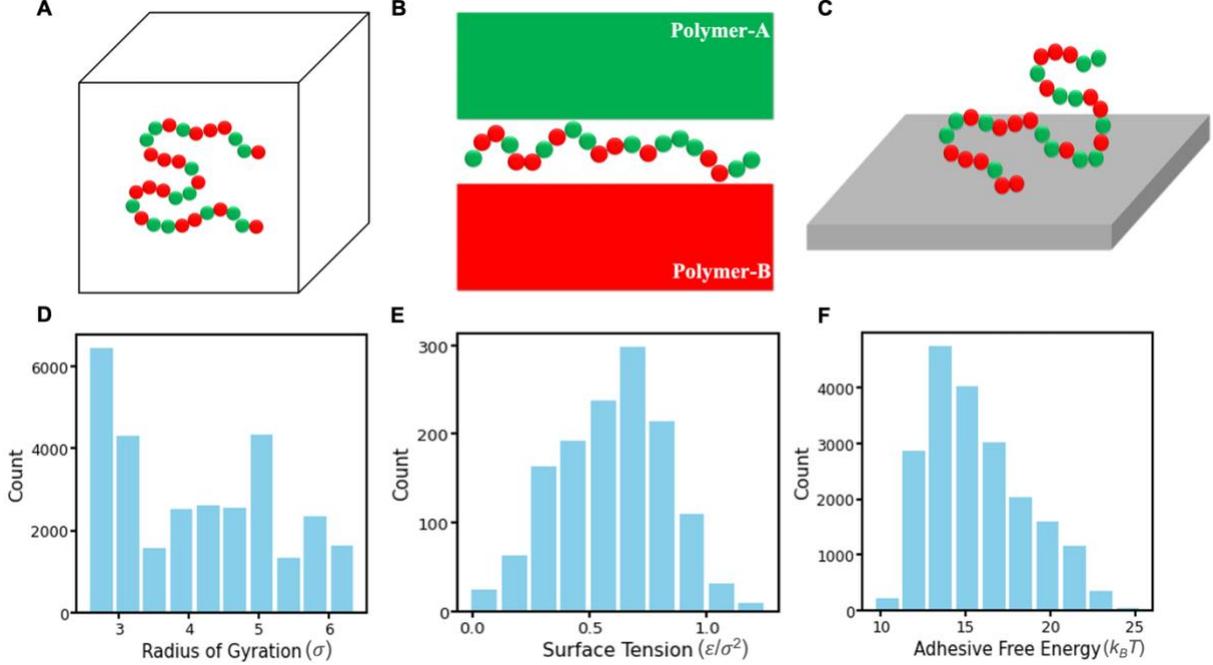

*Figure 1: Sequence-property polymer data. The schematic representations of three systems - folding of a polymer chain in an implicit solvent, a copolymer compatibilizer at the interface between two immiscible homopolymers, adsorption of a copolymer on a substrate are shown schematically in A, B and C, respectively. The corresponding histograms of the available data for the three cases are shown in D, E and F, respectively. A reduced unit is used in all three studies, wherein σ and ε are the unit of length and energy, respectively. Also, $k_B$ and T are the Boltzmann constant and temperature of a system, respectively.*

## III. Supervised Deep Neural Network (DNN)

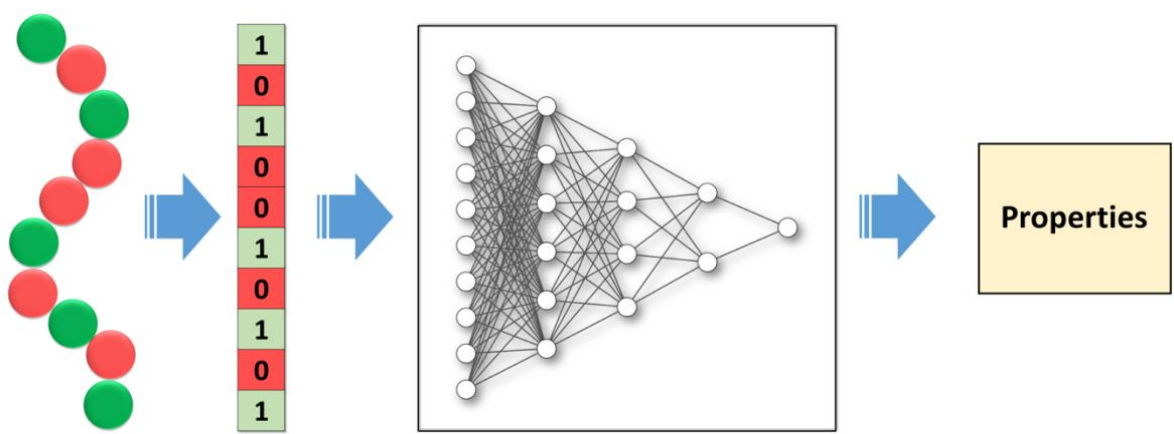

*Figure 2: Schematic representation of the deep learning method. A polymer is mapped to a machine-readable fingerprint, viz., a binary number. The fingerprint serves as an input to the model and a polymer property is the output. During the training, the weights between the nodes of the deep neural network are optimized. Hyperparameters – the number of hidden layers and nodes in all the hidden layers are decided using a linear expansion algorithm. The dimension of the input layer is the same as that of the polymer fingerprint.*

We build a fully connected feedforward backpropagation DNN for predicting sequence-defined properties of polymers, as schematically shown in Figure 2. The DNN consists of many intermediate layers of perceptions, and these layers aid in polymer feature learning. The polymer sequences are encoded into a fingerprint that the model can read and understand. We transform the polymer sequence data into an array consisting of *0*s and *1*s, where *0* represents one type of moiety in the binary copolymer, while *1* represents the second type of moiety. This binary representation is fed to the DNN via the input layer. Therefore, the number of nodes in the input layer is equal to the polymer chain length. The output layer of the DNN consists of one node, which represents the target property of the copolymer. The number of intermediate layers, which are also known as hidden layers, and the number of nodes in each of the intermediate layer is varied to improve the performance of the model. We also use the standard dropout regularization procedure to avoid overfitting and improve the efficacy of the model.[45] During the network training, randomly selected nodes are dropped out. We use a dropout rate of 0.2 in this work. All the nodes in the hidden layers are activated using the rectified linear unit (ReLU)[46,47] activation function, which can be written as $f(x) = \max(0, x)$. This is a piecewise linear function that outputs the input directly if it is positive; otherwise, it will output zero. The output layer node is activated using a linear activation function that outputs the property as $y = ax$. Here $x$ is the input, and we choose $a = 1$. A node in the DNN receives the weighted sum of all the signals from all the nodes of its previous layer and outputs the activated signal, which is fed to all the nodes of its next layer. This feedforward process produces an output in the output layer node, which is the prediction of the model. We define mean square error (MSE) as the loss function of the DNN model, which represents the difference between the output of the model and the true value. Therefore, the loss function is written as $L(y, \hat{y}) = \frac{1}{N}\sum_{i=0}^{N}(y_i - \hat{y}_i)^2$. Here, $y$ and $\hat{y}$ are the actual and predicted quantities in the output layer node, respectively. The loss in the output layer node is backpropagated to estimate errors in all the compute nodes in the previous layers.[48] During the training, the weights are optimized using the Adam optimizer[49] to reduce the loss in all the compute nodes in the network. The training is stopped when the loss function reaches a plateau, and a certain minimum value of the loss is obtained, which doesn't improve in further cycles. Subsequently, the trained DNN is used to predict the properties of unseen polymer sequences. We use the open-sourced Keras application programming interface (API) to build the model.[50] This procedure of DNN model building is used in all the subsequent sections of the article.

```
Algorithm: DNN Hyperparameter Optimization
1:  Num_Layer = 0                              # Number of Hidden layers
2:  Max_Layer = 1000, Max_Node = 10000         # Initializing with High values
3:  Loss_Layer = 0                             # Initializing with Low value
4:  WHILE ( Num_Layer < Max_Layer ) DO
5:          Num_Layer ++                       # Adding a new Hidden Layer
6:          Num_Node = 0                       # Starting with zero nodes in the new layer
7:          Loss = High Initial Value          # Initializing with High Value
8:          WHILE ( Num_Node < Max_Node ) DO
9:                  Num_Node += 10
10:                 Train the Model for certain epochs
11:                 New_Loss                   # Calculating loss after adding extra 10 nodes
12:                 IF ( New_Loss > Loss ) DO
13:                         BREAK              # Breaking Inner loop
14:                 end IF
15:                 Loss = New_Loss
16:         end WHILE
17:         IF ( Loss_Layer > Loss ) DO
18:                 BREAK                      # Breaking Outer loop
19:         ELSE DO
20:                 Loss_Layer = Loss          # Would lead to adding another layer
21:         end IF
22: end WHILE
```

*Figure 3: A pseudo code for topology optimization of a DNN. The number of nodes added to the network in each step is a user-defined variable. We consider it to be 10 for all the case studies.*

IV.     DNN Topology Selection

DNN model development deals with many hyperparameters, including DNN topology, optimizer, learning rate, and batch size. Among these hyperparameters, we find that DNN topology selection is more challenging than others. Here our interest is to understand the correlation between a DNN topology and its performance and develop a simple strategy to decide on an efficient DNN topology for a regression task. The input layer and the output layer sizes correspond to the input and output data dimensions, respectively. Therefore, the dimension of these two layers are fixed for a given problem. The variables are the number of hidden layers and the number of nodes in all the hidden layers. There is a very limited understanding of how these two variables – depth and width of a DNN impact the learned representation and its overall performance.[51,52] Here, we study the impact of the depth and width of the DNN model on a polymer sequence problem. We systematically expand the DNN starting from the first layer. We stop the growth of a given layer when the loss function does not improve while adding more nodes and move to the next layer, as shown by the pseudo-code in Figure 3. This layer-by-layer growth is stopped when the loss function does not improve by adding any additional layer. This simple approach leads to a DNN that shows the best performance for a given problem. A representative case study is shown in Figure 4 for a DNN model that predicts the sequence-defined surface tension of a polymer blend. As indicated in Figure 4A, the loss function decreases as the number of nodes increases for a DNN with one hidden layer. We stop growing the size of the hidden layer when the loss function does not improve significantly while adding more nodes. Subsequently, we build DNN with two hidden layers, where the dimension of the first-hidden layer is the same as the last one, and progressively increase the nodes in the second layer. It shows further improvement. We continue the process up to a DNN with six layers. As shown in Figure 4, the best performance is achieved for a DNN with three hidden layers. The performance of DNNs with more than three hidden layers tends to degrade. The optimized topology for this surface tension model is 20-250-300-50-1. The coefficient of determination ($R^2$) of this model is 0.95 for unseen data.

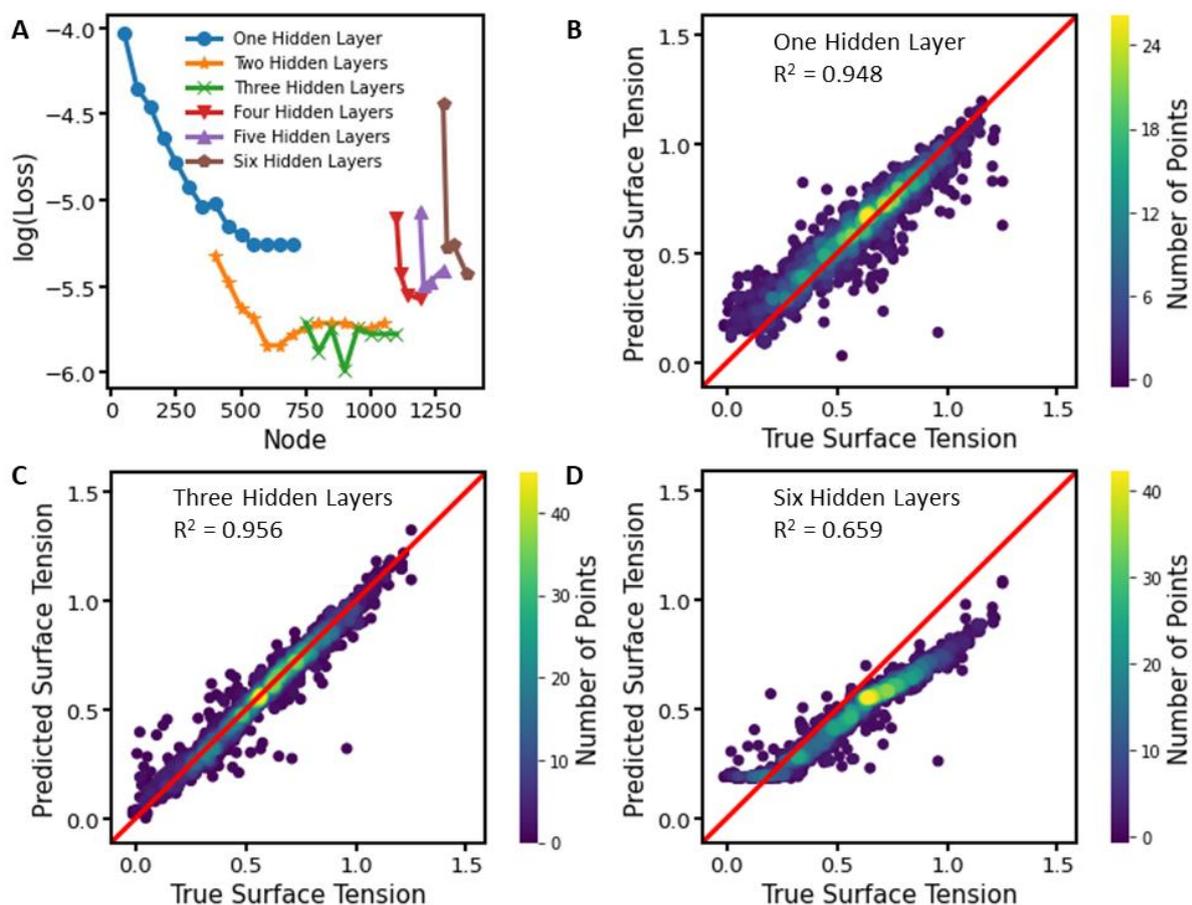

*Figure 4: The performance of a DNN as a function of its topology. The number of inputs to the DNN is 20 representing the sequence of moieties in a binary copolymer compatibilizer of length 20 and the output represents the surface tension of the system. The lowest value of the loss function is shown as the cumulative number of nodes in a DNN with varying topology is shown in A. The performance of best DNN models with one, three, six hidden layers are shown in B, C and D, respectively. The surface tension values are in a reduced unit.*

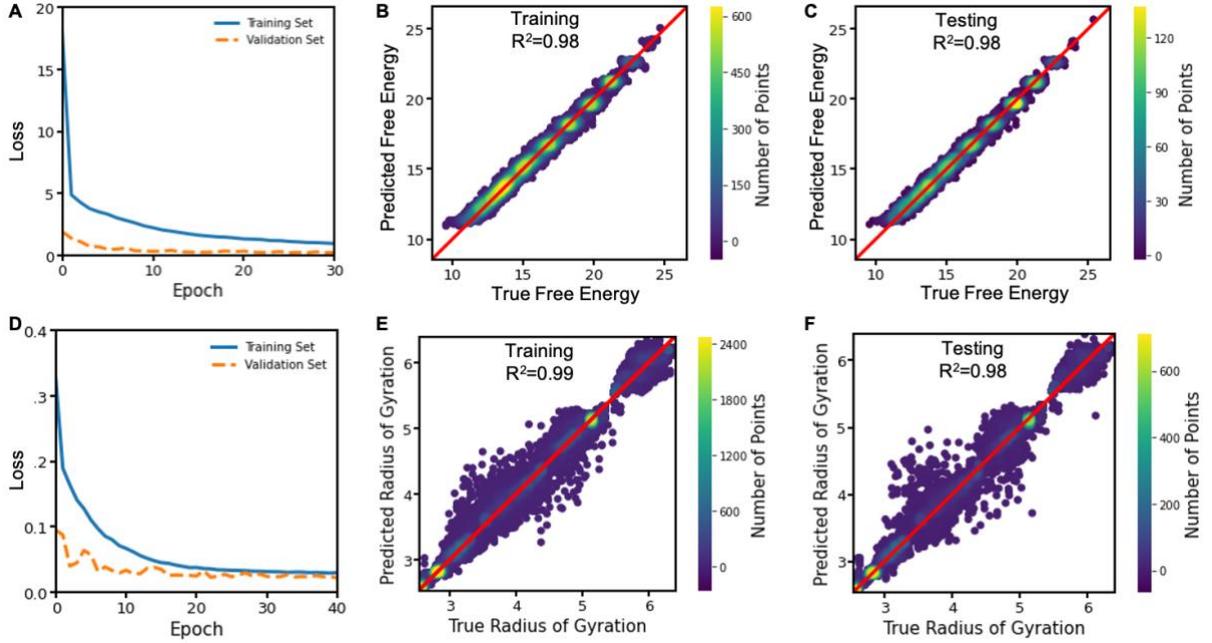

*Figure 5: Performance of topology-optimized DNN models. Panels - A, B and C correspond to the adhesive free energy predictions. Panels - D, E and F correspond to the radius of gyration predictions. The training and test data consist of 80% and 20% of the total data. The validation set consists of 10% of the training data. The loss function during the training is shown in A and D for the free energy model and radius of gyration model, respectively. The model performance on training and test data is shown in (B, C) and (E, F) for free energy model and radius of gyration model, respectively. Here the radius of gyration and free energy values are in a reduced unit.*

The above layer-by-layer growth of a DNN appears to be a simpler and more efficient strategy for topology selection. This does not require a computationally expensive optimization algorithm. We further employ this approach to identify the best DNN topology for adhesion-free energy and radius of gyration prediction, as shown in Figure 5. The optimized topology for the adhesive free energy model is 20-100-50-1, and that for the radius of gyration model is 100-200-100 -1. We note that our layer-by-layer growth strategy leads to a DNN where the first few layers' dimensions are higher than their previous layer. We also observe that such a "wide and then narrow" topology performs better than conventional narrow topologies. To understand this characteristic further, we compare the performance of two topologies - one with "wide and then narrow" architecture and the second one with a narrow architecture, as shown in Figure 6 for the surface tension surrogate model. We consider a case wherein both the topologies contain 350 nodes in their hidden layers. The architecture of the "wide and then narrow" DNN is 20-200-100-50-1 while the narrow DNN architecture is 20-20-20-20-20-20-20-20-20-19-18-17-16-15-14-13-12-11-10-9-8-7-6-5-4-3-2-1-1. As a consequence, the total adjustable parameters for the two cases are 29401 and 6212, respectively. The number of parameters can grow rapidly by increasing the width of the layers without making them deep by increasing more hidden layers of perception. This rearrangement of the nodes leads to a significant change in the number of model parameters. As shown in Figure 6F, the narrow network completely fails to learn the correlation. We further test a large number of narrow

DNNs with varying depth and width and identify 20-20-15-1 as the best-performing narrow topology with a $R^2$ score of 0.54 for test data set. However, the best narrow DNN is still inferior to the "wide and then narrow" topology, which is identified by our layer-by-layer expansion algorithm. We infer that the wide network with two/three intermediate layers performs better than a narrow DNN with a large number of intermediate layers for these polymer problems. Recent theoretical work suggests that such a "wide and then narrow" DNN can provide accurate and compact topology.[53]

## V. Extrapolation using Deep Neural Network Model

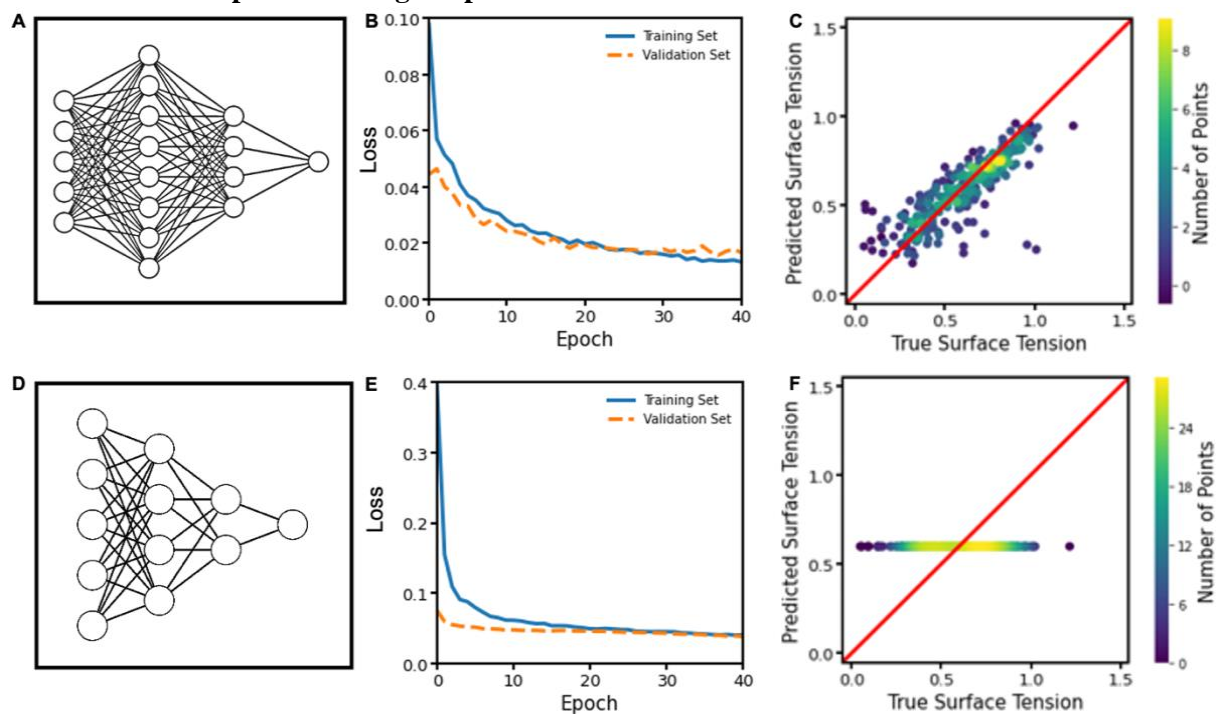

*Figure 6: Impact of DNN topology on its performance. The panels – A, B and C correspond to a 'wide and then narrow' topology. The panels – D, E and F correspond to a narrow topology. The figures A and D are schematic representations of "wide and then narrow" and strictly narrow topologies, respectively. The loss function of these two DNNs during the training of surface tension data is shown in B and E, respectively. The model performance on unknown data is shown in C and F for the "wide and then narrow" and strictly narrow topologies, respectively. Here the surface tension values are in a reduced unit.*

ML models are inherently interpolative. However, as the mathematical framework of an ML model is flexible, the general perception is that it can be used to predict a wide range of properties, given that it is built upon a large number of data points. There have been few attempts to use ML models to predict property outside the known range of property. However, its ability to extrapolate is not rigorously tested. Here we curate two subsets of data in such a way that they represent two different ranges of a material's property. One of these two subsets of data is used to build a DNN model, and the performance of the model is tested on the second subset of data, as shown in Figure 7 for two representative cases. We consider a wider range for the test data than the training data. The training and test set consists of 7000 and 13000 data points for the adhesion free energy model. Among the test data, 11000 of them are outside the property range of training data. Likewise, we use 7000 training data points and 22000 test data points for the radius of gyration model. In this case, 20000 test data points fall outside the property range of training data. This partitioning ensures that we test the model's capability for truly outside the range of training data. As shown in Figures 7B and E, both models perform well when the test data are within the range of training data. The performance of the model degrades drastically when the test data is far away from the range of training data. The $R^2$ of the free energy model is 0.96 and 0.77 for the within-the-training-range test data and outside-the-training-range test data, respectively. In the case of the radius of gyration model, $R^2$ is 0.85 and 0.61 for within-the-training-range test data and outside-the-training-range test data, respectively. We also plot the absolute error in prediction for all the test data points as a function of their actual value in Figure 7C and F for the adhesion free energy model and radius of gyration model, respectively.

It clearly shows how systematically and monotonically the model's performance degrades as we move slowly away from the range of training data. We note that the models perform reasonably well for many nearby points of the training data range. Therefore, our analysis suggests that the training-target mismatch is a fundamental problem of ML model development.

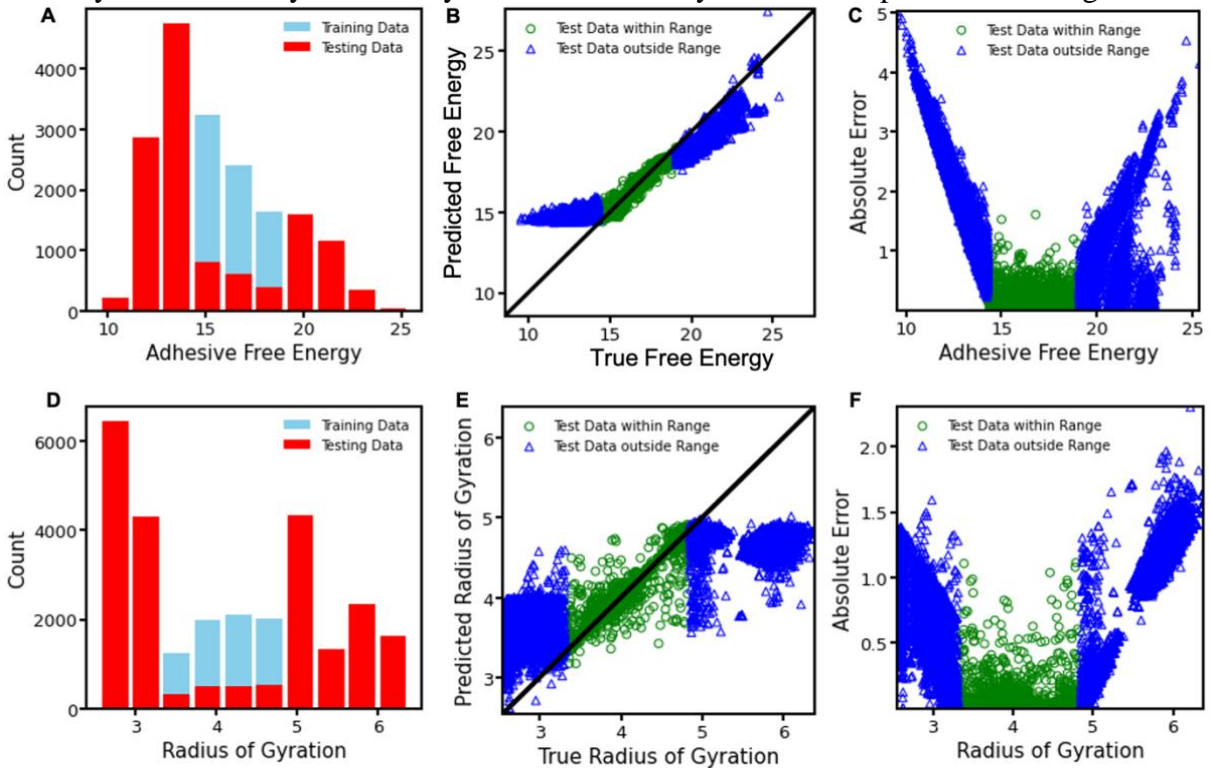

*Figure 7: DNN extrapolation. The panels – A, B and C correspond to the free energy model. The panels -D, E and F correspond to the radius of gyration model. The histograms of training and test data are shown in A and B. The models' performance on test data are shown in B and E. The absolute error in predicted-property of a polymer sequence is plotted against its actual property in C and F for the free energy metamodel and the radius of gyration model, respectively. The radius of gyration and free energy values are in a reduced unit.*

## VI. Intelligent Data Selection via Unsupervised Deep Learning

As discussed in the previous section, the success of an ML model primarily depends on the diversity of the training data. Randomly selected data may fall in a narrow region of the configurational space of a material. In such cases, the machine learns the properties of the material for those regions of the configurational space that are well represented in the training set and struggles to learn the properties well outside those regions. Therefore, the success of an ML model relies on capturing the entire range of properties in the training data set. Moreover, one can ask how many data points from a region of the configuration space are enough to learn the property of all the candidate structures that belong to the same region of configuration space. There are no straightforward approaches to tackling these issues. To address these issues, we establish a continuous one-dimensional (1D) representation of the conformational space of a binary copolymer using unsupervised deep learning. The idea is to estimate the range of conformational space a priori. This 1D representation of polymers helps us to select training data uniformly across the entire range of configurational space. It also allows us to reduce the number of points in the training set without compromising its performance. We build a convolutional deep autoencoder for this purpose.

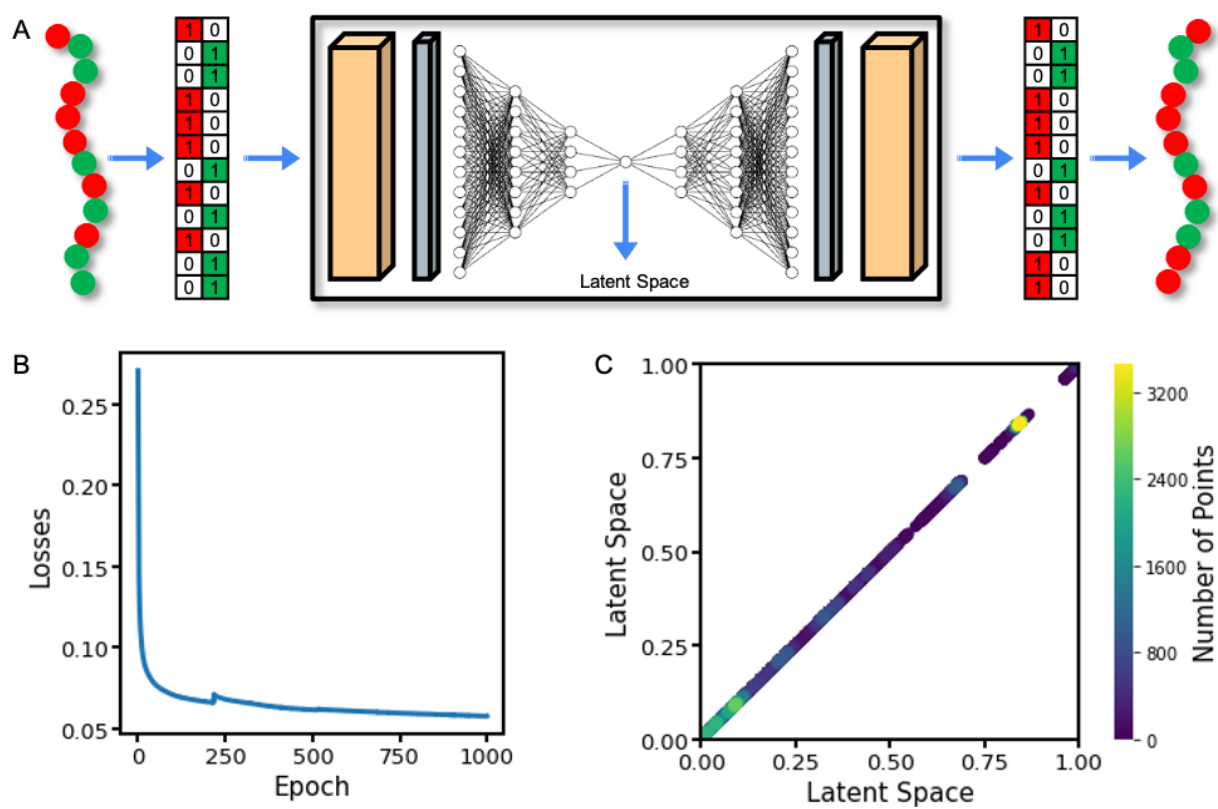

*Figure 8: Deep convolutional autoencoder. The workflow of the unsupervised machine learning model development is shown schematically in A. One-hot encoded fingerprints of a copolymer is fed to the network which also serve as the output of the network. The loss function of the network during the training is shown in B. The one-dimensional latent variable for 30000 sequences is plotted in a two-dimensional plot for visual inspection of the range and continuity of the latent variable in C.*

The autoencoder combines a convolutional neural network (CNN) and a DNN into a single unsupervised ML pipeline, as schematically shown in Figure 8A. CNNs are a special kind of neural network that employs a three-dimensional neural pattern inspired by the visual cortex of animals. They are primarily used for image recognition, object detection, and natural language processing.[54,55] Here, we use a CNN for the efficient construction of a 1D latent space of the sequence space of a polymer. As shown in Figure 8A, a DNN is sandwiched between convolutional and deconvolutional layers in such a way that the input signal is compressed to a 1D variable in the middle of the network, and it remaps back to the actual signal at the output of the network. Thus, the convolutional autoencoder has an encoder and a decoder part. The encoder consists of two convolutional layers followed by a flatten layer, and then we arrange six deep layers with a successively decreasing number of nodes. The topology of the encoder part of the DNN is 100-60-30-10-3-1. A convolutional layer applies a convolution filter to the input data to detect features and pass it to the next layer. The filters in the successive convolution layers are 4 and 2, respectively. We use a kernel of size (2, 2) for the convolution operations. The flatten layer is used to transform the output obtained from the convolutional layers into a 1D array so that it can be fed to the successive deep layers of the encoder. The decoder part of the DNN is a mirror image of the DNN encoder. Therefore, it consists of six deep layers with an increasing number of nodes. The architecture of the decoder part of the DNN is 1-3-10-30-60-100. We add a reshape layer to transform the output from the last layer of the DNN to a 3D matrix that can be deconvolutionized to construct the output of the ML pipeline. Two deconvolution layers are used for this purpose. The filters in the successive deconvolution layers are 2 and 4, respectively. A kernel of size (2, 2) is used for the deconvolution operations. Rectified linear unit (ReLu) activation function is utilized to activate the signals as they pass through all nodes from the input to the output of the ML pipeline, except the latent space node.[46,47] The input and output nodes of the latent space are activated by a linear activation function. A detailed description of the convolution operation, flattening, reshaping, and deconvolution operation can be seen elsewhere.[56,57] We use the one-hot encoded image of a polymer sequence as the input and output of the ML pipeline. A copolymer of chain length 100 is represented as a 100x2 matrix, as shown in Figure 8A. The columns of the fingerprint matrix represent the two types of moieties that construct the copolymer. The rows correspond to the monomers of the copolymer chain starting from one end. In this matrix, 0 and 1 represent the presence and absence of a moiety, respectively, at a given location along the polymer chain. This binary matrix of dimension 100x2 is converted to a binary image. All the copolymers of the data set are converted to binary images for training the model. The Adam optimizer is used during the training to optimize the weights of the model.[49] We use Keras API for developing the model.[50]

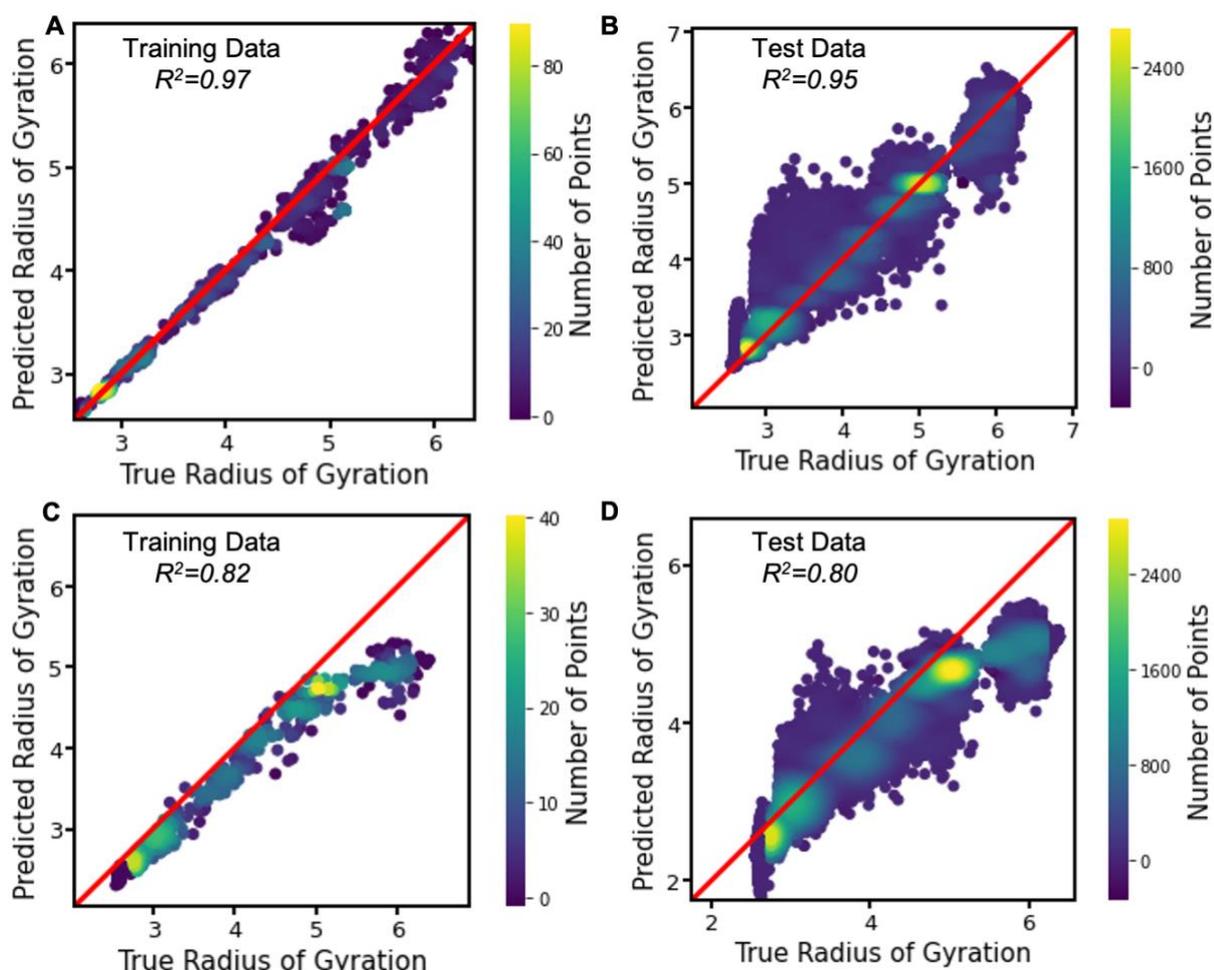

*Figure 9: Performance of a deep learning radius of gyration model for a curated training data set and randomly selected training data is shown in (A,B) and (C,D) respectively. Total number of training data point is 500 for both cases. The model is tested for a data set of 29000 points. The radius of gyration values are in a reduced unit.*

We use 30000 data points for training the convolutional autoencoder. The autoencoder is compiled using the Adam optimizer with a learning rate of 0.0001, with mean square error as the loss function. The training is done for 1000 epochs. The loss function during the training of the convolutional autoencoder is shown in Figure 8B. It suggests that the autoencoder is able to compress the sequences to a lower dimensional representation very accurately. We plot the latent space representation of all the 30000 sequences in Figure 8C. All the points are densely populated within a range of the latent variable. We now build two DNN models. First, we build a DNN by uniformly selecting 500 data points across the latent space. These 500 points are equidistant in the latent space. The second model is built by randomly selecting 500 data points from the database. We test the model performance on the remaining data (~29000). The topology of the DNN is 100-300-50-1 for both cases. As shown in Figure 9, the DNN model with intelligent data selection performs better than the one built upon randomly selected data points. This method does not require properties of materials for deciding which candidate structures to be used for training. Therefore, it can potentially be used to guide data generation in the absence of data. It can also ensure a minimum number of expensive physics-based property calculations for an ML model development. Moreover, the traditional random split of data for training and testing of an ML model can be replaced by this rational approach that ensures diversity and enhances the representation of candidate structures in a training set from the entire configuration space of a molecule.

VII. Conclusions

Machine learning and artificial intelligence bring huge promises to solve complex material science and engineering problems. The success of these ML and AI-driven materials research relies on a deeper understanding of these tools, their generalizability, and their explainability. Here we present a holistic view of deep learning model development, its advantage, and its limitation for polymer science problems. We present a systematic approach to building a minimal deep-learning architecture for modeling sequence-property correlations of polymers. We demonstrate the sensitivity of a DNN's architecture to learn polymer features. We propose that instead of using a computational expensive optimization algorithm such as the Bayesian algorithm or genetic algorithm, one can systematically grow a DNN layer-by-layer to identify an efficient topology that makes good-quality predictions. This layer-by-layer expansion of a neural network topology is computationally efficient and performs better than intuitive and random architectures. We carefully analyze the transferability of deep learning models across the sequence space of a polymer. We show that the models' performance tends to decline for points in a different property range than the training data. To address this transferability problem, we propose a new method for training data selection, wherein candidate structures are mapped to a lower dimensional latent space to better understand the conformational space of a polymer. We then select data representing a polymer's entire conformational space in its latent space representation and build a deep-learning model using these data points. Our analysis clearly suggests that such an rational data selection is better than a random data section for building machine learning models. As this data selection method does not require the property, it can be used to select a small number of candidate structures among a large number of possibilities for physics-based property calculation and subsequently use them for model development. Such a transferable model will be of great use in the search for optimal materials. We note that an array of binary numbers is used as the polymer fingerprint for the predictive model development, while a binary image is used as a polymer fingerprint for the latent space creation. The choice of the binary image as a polymer fingerprint is to take advantage of convolutional neural network for feature extraction. Although we use the data selection algorithm and topology selection algorithm for sequence-property models of polymers, we expect that these algorithms are extensible for other materials systems. The work will have important implications in feature learning and machine learning model development for polymers and other materials.

**Acknowledgements.**

The work is made possible by financial support from SERB, DST, Gov of India through a start-up research grant (SRG/2020/001045) and National Supercomputing Mission's research grant (DST/NSM/R&D_HPC_Applications/2021/40). This research used resources of the Argonne Leadership Computing Facility, which is a DOE Office of Science User Facility supported under Contract DE-AC02-06CH11357. We also used the computational facility of the Center for Nanoscience Materials. Use of the Center for Nanoscale Materials, an Office of Science user facility, was supported by the U.S. Department of Energy, Office of Science, Office of Basic Energy Sciences, under Contract No. DE-AC02-06CH11357.We acknowledge the use of the computing resources at HPCE, IIT Madras.

**References.**